\documentclass[a4paper,final]{appolb}
\usepackage{graphicx}
\usepackage[colorlinks=true,linktocpage=true,linkcolor=blue,citecolor=blue]{hyperref}
\usepackage[usenames,dvipsnames]{color}
\usepackage{amsmath, amssymb,oldgerm}
\usepackage{multirow}
\usepackage{color}
\usepackage[normalem]{ulem}  
\usepackage{braket}
\usepackage{slashed}
\usepackage[mathscr]{eucal}
\usepackage{tikz}
\usepackage{pgfplots}				
\usepackage{caption}
\usepackage{placeins}
\usepackage{pifont}
\usepackage{xcolor}

\newcommand{\mC}{\mathfrak{C}}

\newcommand{\ms}{\mathfrak{s}}

\newcommand{\beq}{\begin{equation}}
\newcommand{\eeq}{\end{equation}}

\newcommand{\eq}{Eq.~}

\newcommand{\mfp}{\text{mfp}}
\newcommand{\hydro}{\text{hydro}}

\definecolor{goethe-blau}{cmyk}{1.0,0.2,0.0,0.4}
\definecolor{hellgrau}{cmyk}{0.04,0.04,0.05,0.02}
\definecolor{sandgrau}{cmyk}{0.12,0.09,0.13,0.0}
\definecolor{dunkelgrau}{cmyk}{0.25,0.25,0.30,0.75}
\definecolor{emo-rot}{cmyk}{0.04,1.0,0.8,0.07}
\definecolor{green}{cmyk}{0.08,1.0,0.3,0.36}
\definecolor{senfgelb}{cmyk}{0.01,0.25,1.0,0.05}
\definecolor{gruen}{cmyk}{0.62,0.4,0.87,0.09}
\definecolor{magenta}{cmyk}{0.08,0.86,0.12,0.12}
\definecolor{orange}{cmyk}{0.0,0.7,1.0,0.04}
\definecolor{sonnengelb}{cmyk}{0.0,0.12,0.95,0.0}
\definecolor{helles-gruen}{cmyk}{0.4,0.17,0.81,0.07}
\definecolor{lichtblau}{cmyk}{0.8,0.0,0.06,0.04}
\definecolor{green}{rgb}{0,0.6,0}

\begin{document}
\title{New developments in relativistic hydrodynamics%
\thanks{Presented at Quark Matter 2022}%
}
\author{Nora Weickgenannt
\address{Institute for Theoretical Physics, Goethe University,
Max-von-Laue-Str.\ 1, D-60438 Frankfurt am Main, Germany}
 }
\maketitle
\begin{abstract}
We review recent progress in relativistic hydrodynamics, discussing causal and stable first-order hydrodynamics, known as BDNK theories, hydrodynamic attractors, as well as hydrodynamics near the chiral critical point and spin hydrodynamics.
\end{abstract}
  
\textbf{Introduction --- }
Relativistic hydrodynamics is a well-established theory with a long history and a wide range of applications. In particular, it turned out to extremely successfully describe the so-called quark-gluon plasma created in relativistic heavy-ion collisions~\cite{Heinz:2013th,Gale:2013da}. Hydrodynamic behavior is characterized by the property that a system can be effectively described without full knowledge of its microscopic properties by fundamental principles such as conservation laws. Standard hydrodynamics is based on the conservation equations for the charge current $N^\mu$ and the energy-momentum tensor $T^{\mu\nu}$,
\begin{equation}
    \partial_\mu N^\mu=0\; , \hspace{1cm} \partial_\mu T^{\mu\nu}=0\; .
\end{equation}
In traditional approaches, hydrodynamics is regarded as the classical, long-time and long-wavelength limit of the underlying fundamental quantum theory. In particular, this means that the system is assumed to be near local equilibrium, implying that the so-called Knudsen number Kn$\equiv l_{\text{mfp}}/l_{\text{hydro}}$, where $l_\mfp$ is the mean free path and $l_\hydro$ is a scale characterizing inverse gradients of hydrodynamic fields, is sufficiently small. However, recent developments suggest that hydrodynamics may give a useful description of a system even beyond the mentioned limits, extending hydrodynamic frameworks in order to include far-from-equilibrium dynamics~\cite{Heller:2015dha,Blaizot:2019scw,Kurkela:2019set,Giacalone:2019ldn,Almaalol:2020rnu,Blaizot:2020gql,Ambrus:2021sjg,Blaizot:2021cdv,Chattopadhyay:2021ive,Jaiswal:2021uvv}, short-wavelength phenomena like chiral symmetry breaking~\cite{Stephanov:2017ghc,An:2019csj,Grossi:2020ezz,Grossi:2021gqi,Florio:2021jlx} or quantum phenomena like spin~\cite{Florkowski:2017ruc,Montenegro:2020paq,Bhadury:2020cop,Weickgenannt:2020aaf,Speranza:2020ilk,Fukushima:2020ucl,Bhadury:2020puc,Li:2020eon,Hongo:2021ona,Florkowski:2021wvk,Wang:2021wqq,Hongo:2022izs,Ambrus:2022yzz,Singh:2022ltu,Daher:2022xon,Weickgenannt:2022zxs,Weickgenannt:2022jes,Bhadury:2022qxd}. Furthermore, there has been important progress in the past years in formulating causal and stable first-order hydrodynamics, known as Bemfica-Disconzi-Noronha-Kovtun (BDNK) theories~\cite{Bemfica:2017wps,Bemfica:2019knx,Kovtun:2019hdm,Hoult:2020eho,Bemfica:2020zjp,Speranza:2021bxf}. 

\textbf{First-order hydrodynamics --- }
In many contexts hydrodynamics can be derived by an expansion in Knudsen numbers around local equilibrium. Up to first order, such an expansion yields the so-called Navier-Stokes theory. However, in the relativistic case Navier-Stokes equations of motion violate causality~\cite{hiscock1983stability}, leading to numerical instability, and are therefore not applicable in relativistic systems. 
Recently, it was found in Refs.~\cite{Bemfica:2017wps,Bemfica:2019knx,Kovtun:2019hdm} that the acausality and instability of relativistic first-order hydrodynamics can be cured, leading to the so-called BDNK formulation of first-order hydrodynamics. Consider the decomposition of the energy-momentum tensor with respect to the fluid velocity $u^\mu$,
\begin{equation}
    T^{\mu\nu}=(\epsilon_0+\mathcal{A}) u^\mu u^\nu -\Delta^{\mu\nu}(P_0+\Pi)+u^{\mu} h^{\nu}+u^{\nu} h^{\mu}+\pi^{\mu\nu},
\end{equation}
with $\Delta^{\mu\nu}\equiv g^{\mu\nu}- u^\mu u^\nu$, $\epsilon_0$ and $\mathcal{A}$ being the equilibrium and nonequilibrium energy density, respectively, $P_0$ the thermodynamic pressure, $\Pi$ the bulk viscous pressure, $h^\mu$ the heat current, and $\pi^{\mu\nu}$ shear-stress tensor. 
Up to first order in Knudsen numbers, the dissipative components of $T^{\mu\nu}$ can be expressed as gradients of temperature and fluid velocity, such that the conservation of energy-momentum in principle yields a closed system of differential equations to determine these quantities, given that a frame-choice condition as well as so-called matching conditions are supplemented. In Navier-Stokes theory, one commonly chooses the so-called Landau frame in order to define the fluid velocity, i.e., one sets $h^\mu\equiv0$ and $\mathcal{A}\equiv0$. In this case, the system of equations of motion is acausal and unstable. However, the explicit form of these equations depends on the frame choice as well as on the matching, and therefore these conditions can be chosen in a way which ensures causality, stability and well-posedness.  
This procedure has successfully been applied to derive causal, stable and well-posed first-order hydrodynamics in Refs.~\cite{Bemfica:2017wps,Bemfica:2019knx,Kovtun:2019hdm,Hoult:2020eho,Bemfica:2020zjp}. Furthermore, in Refs.~\cite{Noronha:2021syv,Rocha:2021lze} also transient hydrodynamics with a general frame choice has been studied.

\textbf{Hydrodynamic attractors --- }
A maybe surprising feature of hydrodynamics is its ability to describe certain systems which are not close to local equilibrium, in other words, systems with large Knudsen number. In this context, the existence of so-called attractor solutions~\cite{Heller:2015dha}, which determine the early- as well as late-time behavior of these systems, turned out to be of significant importance. 
In the following, we will review as an example studies on attractors and fixed points in kinetic theory and hydrodynamics for boost invariant systems both in the conformal limit \cite{Blaizot:2019scw,Blaizot:2020gql,Blaizot:2021cdv} and in the massive case~\cite{Chattopadhyay:2021ive,Jaiswal:2021uvv}. 

While hydrodynamics is characterized by macroscopic equations of motion, kinetic theory describes microscopic dynamics by the Boltzmann equation
\begin{equation}
    p^\mu \partial_\mu f(x,p)=C[f]\; , \label{boltz}
\end{equation}
where $f(x,p)$ is the distribution function and $C[f]$ is the collision term. 
From \eq\eqref{boltz} various types of dissipative hydrodynamic equations of motion can be derived by using different methods, such as an expansion in Knudsen numbers, referred to as Chapman-Enskog expansion, or the method of moments, where certain moments of the distribution function with respect to momentum are treated dynamically, leading to transient fluid dynamics. Examples for the latter theories are Israel-Stewart (IS)~\cite{Israel:1979wp}, Denicol-Niemi-Molnar-Rischke (DNMR)~\cite{Denicol:2012cn}, or anisotropic hydrodynamics (a-hydro)~\cite{Martinez:2010sc,Florkowski:2010cf}.

One way to explore the suitability of these theories to correctly describe the microscopic dynamics is to analyze attractors and fixed points both of the Boltzmann equation and the respective hydrodynamic equations of motion. 
In Refs.~\cite{Blaizot:2019scw,Blaizot:2020gql,Blaizot:2021cdv} the authors derive equations of motion for the moments of the distribution function
from the Boltzmann equation, analyze their fixed points and identify attractor solutions in a conformal boost-invariant system. It is found that for a boost-invariant expanding plasma the transition from the free-streaming to the hydrodynamic regime (i.e., the regime of small Knudsen number) is controlled by an attractor, related to the presence of fixed points in both regimes. Furthermore, already the two-moment truncation of the equations of motion describes the evolution of the system well, since the fixed points are already present in this truncation and slightly modified by the coupling to higher moments. Interestingly, it is found that a modification of the second-order transport coefficients in order to put the fixed points on the right location in the free-streaming regime leads to a formulation of second-order viscous hydrodynamics which very accurately describes the exact solution of the Boltzmann equation already at early times.




		

We now turn to the case of a nonconformal, boost-invariant expanding system considered in Refs.~\cite{Chattopadhyay:2021ive,Jaiswal:2021uvv}. When studying the transition from the free-streaming to the hydrodynamic regime, it is found that, in contrast to the massless case, no early-time attractor for the pressure $\Pi$ nor for the shear stress $\pi$ (which is in Bjorken symmetry the only nonzero  component of the shear-stress tensor $\pi^{\mu\nu}$) exists. On the other hand, defining the longitudinal pressure $\mathcal{P}_L\equiv P_0+\Pi-\pi$, it turns out that the solution $\mathcal{P}_L=0$ acts as an attractive fixed point for the equation of motion of the distribution function, and therefore an early-time attractor solution for the longitudinal pressure can be identified. Furthermore, a comparison of hydrodynamic frameworks shows that, in contrast to hydrodynamics obtained by a Chapman-Enskog expansion,
anisotropic hydrodynamics features the correct locations of the fixed points and therefore describes the longitudinal pressure well even at early times.

\textbf{Hydrodynamics with chiral symmetry breaking --- }
One of the main purposes of studying relativistic heavy-ion collisions is to explore the phase diagram of quantum chromodynamics (QCD). In order to make use of a hydrodynamic approach to describe the chiral phase transition, the hydrodynamic model has to be extended in order to include additional modes. A well-known example for such an extended hydrodynamic theory is hydro+~\cite{Stephanov:2017ghc}, where fluctuations near the chiral critical point are described by additional slow modes. While this theory first proposed in Ref.~\cite{Stephanov:2017ghc} originally takes into account the slowest critical mode and is valid up to a regime where $k\sim\xi^{-2}$ with $k$ the wave number and $\xi$ the correlation length, it has recently been extended to a regime with $k\sim\xi^{-1}$, i.e., to be valid closer to the chiral critical point, where the correlation length is larger~\cite{An:2019csj}. Furthermore, a freeze-out procedure to convert the chiral critical fluctuations in hydro+ into into measurable cumulants of hadron multiplicities has been developed in Ref.~\cite{Pradeep:2022mkf}.

Another approach to hydrodynamics with chiral symmetry breaking has been proposed in Refs.~\cite{Grossi:2020ezz,Grossi:2021gqi}. These works are based on the idea that below the critical temperature, when a chiral condensate is formed, the chiral symmetry is spontaneously broken, leading to the appearance of Goldstone modes, which are treated as new hydrodynamic modes. The resulting theory is then analogous to a nonabelian superfluid. Since the chiral symmetry is explicitly broken by the finite quark mass, the Goldstone modes become irrelevant on long distances $\ell\gg m_\pi^{-1}$, where $m_\pi$ is the pion mass, and ordinary hydrodynamics is recovered. On the other hand, considering wavelengths of the order $m_\pi^{-1}$, superfluid modes become relevant and result in corrections to the usual transport coefficients. These corrections were calculated in Ref.~\cite{Grossi:2021gqi} starting from an effective action 
and computing current-current and stress-stress correlation functions through linear response theory [see also Ref.~\cite{Florio:2021jlx} for a computation with real-time simulations on the lattice.]

\textbf{Spin hydrodynamics --- }
Recent progress in developing a theory of relativistic spin hydrodynamics is mainly motivated by polarization measurements of Lambda hyperons in noncentral heavy-ion collisions, where the global rotation of the system generates polarization through the conversion of orbital angular momentum into spin~\cite{STAR:2017ckg}. 
While hydrodynamic calculations in local equilibrium assuming spin to be determined by thermal vorticity were able to predict measurements of the global Lambda polarization~\cite{Karpenko:2016jyx}, the same models failed to describe the momentum dependence of the local polarization~\cite{Becattini:2017gcx,STAR:2019erd}. Recently, there has been promising progress towards resolving this puzzle by including contributions from the thermal shear tensor to the polarization in local equilibrium~\cite{Liu:2021uhn,Fu:2021pok,Becattini:2021iol,Becattini:2021suc}. On the other hand, quantitative results for dissipative contributions to the polarization have not obtained up to now. 

The basic idea in spin hydrodynamics is to promote the spin tensor $S^{\lambda,\mu\nu}$, which is part of the total angular-momentum tensor tensor $J^{\lambda,\mu\nu}=x^\mu T^{\lambda\nu}-x^\nu T^{\lambda\mu}+\hbar S^{\lambda,\mu\nu}$, to an additional hydrodynamic variable~\cite{Florkowski:2017ruc}. Its equation of motion is obtained from the conservation of total angular momentum,
\begin{equation}
    \hbar \partial_\lambda S^{\lambda,\mu\nu}=T^{\nu\mu}-T^{\mu\nu}\; .
\end{equation}
In a relativistic theory, the definition of the spin tensor is not unique, but is subject to the so-called pseudo-gauge freedom~\cite{Hehl:1976vr}. While in general the energy-momentum tensor is not symmetric and therefore, the spin tensor is not conserved, it is possible to find pseudo-gauges with conserved spin tensor for free fields, such as the Hilgevoord-Wouthuysen (HW) and de-Groot-van-Leeuwen-van-Weert (GLW) pseudo-gauges~\cite{Speranza:2020ilk}. 

In order to derive dissipative corrections to the spin tensor, a useful starting point is kinetic theory with spin, where the phase-space is enlarged by an additional variable $\ms^\mu$, i. e., the Boltzmann equation is of the form~\cite{Florkowski:2018fap,Weickgenannt:2020aaf,Weickgenannt:2021cuo}
\begin{equation}
    p\cdot \partial f(x,p,\ms) = \mC[f]\;. \label{spinkin}
\end{equation}
In Refs.~\cite{Bhadury:2020cop,Bhadury:2020puc}, the authors make use of a Chapman-Enskog-like expansion of the distribution function with a Boltzmann equation of the form \eqref{spinkin} with the collision term $\mC$ modeled by a relaxation-time approximation to obtain the first-order dissipative corrections to the GLW spin tensor. On the other hand, in Ref.~\cite{Weickgenannt:2022zxs} the method of moments is applied to derive second-order dissipative spin hydrodynamics from a Boltzmann equation of the form \eqref{spinkin} with a nonlocal collision term 
which is responsible for the conversion between orbital angular momentum and spin~\cite{Weickgenannt:2020aaf}. 
In this framework the spin moments corresponding to the dissipative components of the HW spin tensor 
are treated as new dynamical variables of the theory. 
It should be noted that the local collision term determines the relaxation times of the spin moments, while the nonlocal collision term contributes to the Navier-Stokes limit. 
The relaxation times of the spin moments are slightly smaller than those of other dissipative quantities, but of the same order of magnitude. For this reason, it makes sense to treat spin as a dynamical degree of freedom in second-order
dissipative hydrodynamics.

A different way to obtain an explicit form of the spin tensor is through variation of the QCD action  and linear response theory in a general relativistic framework~\cite{Hongo:2021ona}. In order to couple spin and gravity, one considers a torsionful background. The spin tensor is then given by the variation of the action with respect to the spin connection, corresponding to the so-called canonical pseudo-gauge with nonsymmetric energy-momentum tensor. Using linear response theory, dispersion relations for the spin modes are obtained in Ref.~\cite{Hongo:2021ona}, which feature nonhydrodynamic spin modes due to the fact that the spin tensor is not conserved. 
In the considered regime with $T\ll m$, it is found that the spin relaxation rate $\Gamma_s$ is much smaller than the relaxation rate $\Gamma$ of the other nonhydrodynamic modes, $\Gamma_s\sim (T/m)\Gamma\ll \Gamma$, where $m$ is the mass and $T$ is the temperature. 

 \textbf{Conclusions --- }
 Although relativistic hydrodynamics is an old and well-established theory, many of its aspects are not yet fully understood and active research in different directions is still ongoing. Recently, a new theory of causal and stable first-order hydrodynamics, BDNK theory, was developed. Furthermore, new directions of relativistic hydrodynamics were explored, namely far-from-equilibrium hydrodynamics, hydrodynamics with chiral symmetry breaking, and spin hydrodynamics. 

\textbf{Acknowledgments ---}
I thank the organizers of Quark Matter 2022 for the invitation to give this plenary talk. Furthermore, I thank D.\ H.\ Rischke, E.\ Speranza, and D. Wagner for enlightening discussions and fruitful collaboration, as well as F.\ Becattini, J.-P.\ Blaizot, E.\ Grossi, U.\ Heinz, J. Noronha-Hostler, J.\ Noronha, R.\ Ryblewski, M.\ Spalinski, D. Teaney, and Y.\ Yin for valuable advice. The work of N.W.\ is supported by the
Deutsche Forschungsgemeinschaft (DFG, German Research Foundation)
through the Collaborative Research Center TransRegio
CRC-TR 211 "Strong-interaction matter under extreme conditions" -- project number
315477589 -- TRR 211 and by the State of Hesse within the Research Cluster
ELEMENTS (Project ID 500/10.006).

\bibliographystyle{h-physrev}
\bibliography{biblio_proc}{}

\end{document}